\documentstyle[prl,aps,multicol,epsf]{revtex}

\begin{document}

\title{Thermodynamics and quantum criticality in cuprate superconductors.}

\author{J. Zaanen and B. Hosseinkhani}

\address{Instituut Lorentz for Theoretical Physics, Leiden University, 
POB 9506, 2300 RA Leiden, The Netherlands}

 \date{\today}
\maketitle

\begin{abstract}
We will  present  elementary scaling arguments focussed on
the thermodynamics  in the proximity of the
quantum critical point in the cuprate superconductors.
Extending the analysis centered on the Gr\"uneisen parameter by Rosch, Si and
coworkers to the cuprates, we demonstrate that 
a combination of specific heat- and chemical potential measurements
can reveal the nature of the zero temperature singularity. From the known
specific heat data it follows that the effective number of time dimensions
has to equal the number of space dimensions, while we find a total of
six scaling laws governing the temperature and density dependence
of the chemical potential, revealing directly the coupling constant scaling
dimension.  

\end{abstract}


\begin{multicols}{2}

The possible existence of quantum phase transitions (QPT's) 
in a variety of  condensed matter systems is attracting much
interest\cite{sachdev}. The cuprate high Tc superconductors have played a 
prominent role in this development since it has been suspected
for a long time\cite{sachdev1,dicastro,laughlin,varma} 
that the state realized at the doping where the
superconducting transition temperature is maximal ($x_{opt}$) is
controlled by a continuous QPT. This suspicion is mainly motivated
by the observation of the `wedge' in the doping ($x$)-temperature ($T$)
plane set by the `pseudogap' ($T_{SG} (x)$) and  
'coherence'\cite{campuzano,tagaki} ($T_{coh} (x)$)
crossover temperatures, 
bordering a `quantum-critical' (QC) region
characterized by powerlaw behaviors. 
It is believed that this signals a QPT from a
poorly understood `pseudo-gap' phase at low dopings to a Fermi-liquid
at high dopings. Although direct evidences 
appeared for the presence of scale invariance of the quantum dynamics
 in the QC regime\cite{marel},
it is unclear if this `critical state' is truely critical in the sense 
that it is characterized by universality and hyperscaling\cite{sachdev}. 
Given that
apparently fermionic degrees of freedom are involved, this is from a
theoretical point of view far from obvious because the fermion signs
obscure the analogy with thermal phase transitions\cite{senthil}. 
One would like 
to establish emperically the presence of scaling laws, revealing universality. 
Such evidences are lacking in the cuprates.

Thermodynamics has played a pivotal role in establishing the nature of
the classical critical state. In a recent paper, 
Rosch, Si  and coworkers\cite{roschsi}
showed that the thermodynamic singularity structure of  QPT's has quite
interesting observable consequences.  They argued that in the case of a 
QPT where pressure takes the role of zero-temperature control parameter
(`coupling constant' $r$) the Gr\"uneisen parameter (ratio of thermal expansion
and specific heat $C$) is particularly revealing with regard to the presence
of universality. This was subsequently applied succesfully to the QPT's 
in several heavy fermion intermetallics\cite{grueneisen}. 
Here we will adapt and extend 
their scaling analysis to the particular situation encountered in the 
cuprate superconductors. The electronic specific heat of the cuprates
is known\cite{loram}, and using simple scaling arguments we will argue that its
`normal' appearance (i.e. $C = \gamma T$ with $\gamma$ constant) in the
QC- and overdoped regime has actually a profound consequence: it implies
that the effective number of time dimensions associated with the universality
class ($z$, the dynamical critical exponent) has to be equal to the number 
of space dimensions ($d$).
The  quantum ($T=0$) singularity resides elsewhere: the chemical potential 
$\mu$. We find a large set of scaling relations between 
its temperature dependence  and its density dependence (i.e.,
the inverse electronic compressibility) while it also relates directly to
the doping dependence of the pseudo-gap scale $T_{SG}$.
The chemical potential can be measured in
principle with the required accuracy  and such experiments can decide
if a genuine quantum phase transition is taking place in the cuprates. 

Thermodynamics is of course in the first instance associated with temperature.
A classical phase transition is driven by temperature, but this is
profoundly different for a quantum phase transition. The QPT is driven
by a zero temperature control parameter $r$, and the path
integral formalism shows that temperature takes the role of a finite 
size\cite{sachdev}, as the
compactification radius of the imaginary time dimension $L_{\tau} =
\hbar / (k_B T)$. 
The essence of the Rosch-Si scaling analysis\cite{roschsi}
is that one has to determine the dependence of
the  free energy relative to variations of the coupling constant 
to learn about the quantum singularity. However, standard thermodynamics
associated with variations  of temperature gives additional information of the
finite size scaling variety. Their combination yields a powerful 
phenomenological scaling tool box.

Following Si-Rosch\cite{roschsi}, our analysis rests on a single theoretical 
assumption. 
It is assumed that the QPT is associated with an unstable fixed 
point at zero temperature, reached by tuning a single zero temperature 
variable $y$ such that $r = ( y - y_c ) / y_c$ measures the distance
from the critical point residing at $y_c$. Since temperature $T$ corresponds 
with $L_{\tau}$ it enters
the singular part of the free energy density $F_s$ as a finite size under 
a  scale transformation $x \rightarrow b x$, 
\begin{equation}
F_s ( r, T) = b^{- ( d + z)} F_s ( b^{y_r} r, b^z T )\;,
\label{freenscal}
\end{equation}
where $d$ is the space dimensionality and
$z$ the dynamical exponent,  while hyperscaling is assumed in order to
relate the finite size to the scaling dimension $y_r$ of the coupling constant
($y_r = 1/ \nu$ where $\nu$ is the correlation length exponent).
Eq. (\ref{freenscal}) is equivalent to the following 
scaling forms for the free energy density\cite{roschsi}, 
\begin{eqnarray}
F_s ( r, T) & = & - \rho_0 r^{( d + z ) / y_r} \tilde{f} \left(
 { T \over { T_0 r^{z / y_r} } } \right) \;, 
\nonumber \\
& = & - \rho_0 \left( { T \over {T_0} } \right)^{ ( d + z ) / z}
f  \left(
 { r \over { (T/T_0)^{y_r / z} } } \right) \;,
\label{freeenwork}
\end{eqnarray}
where $\rho_0$ and $T_0$ are non-universal constants, while $f(x)$ and
$\tilde f (x)$ are universal scaling functions. Since there is no
singularity at $r = 0, T > 0$, $f (x \rightarrow 0)
\simeq f (0 ) + x f' (0) + (1/2) x^2 f''(0) + \cdots$  
while $\tilde{f} (x) =
\tilde{f} (0) + g ( x )$ where $g(x)$ describes the low temperature
thermodynamics of the phases to the left- or right side of the QPT.
When the phase is fully gapped $ g (x) \sim e^{- 1/x}$ while
for a massless phase $g (x) = c x^{y_0 +1}$ such that $y_0$ corresponds
with its specific heat exponent ($y_0 =1$ for a Fermi-liquid, and 2
for a `nodal liquid' characterized by d-wave like `Dirac cones').  

We find it convenient to parametrize the exponents in terms of $d$, 
$z$ and the zero-temperature analogue of the specific  heat exponent $\alpha$ 
characterizing a thermal phase transition,
\begin{equation}
\alpha_r = 2 - { {d+z} \over {y_r} } \;.
\label{alphar}
\end{equation}
In analogy with classical criticality, we expect this exponent to be
a fraction of unity. Following Rosch-Si, we will consider the specific heat
$C = -T (\partial^2 F / \partial T^2)$ and the quantity 
$\eta_r =   (\partial^2 F / \partial r \partial T)$, revealing the
dependence of the entropy on the coupling constant. However, 
we will extend the analysis by also including the `coupling
constant susceptibility' $\chi_r =  \partial^2 F / \partial r^2$, being
the quantity which is actually most sensitive to the zero-temperature
singularity.    

From the scaling forms Eq. (\ref{freeenwork}) and the above definitions
it follows  that  the singular parts of various measurable quantities
have the following temperature dependence in the quantum critical state
($r = 0$), 
\begin{eqnarray}
C_{cr} (T, r = 0 ) & = & \rho_0 f ( 0 )  { { (d + z) d } \over {z^2} }  
\left( { T \over {T_0} } \right)^{d/z} \;,
\nonumber \\
\eta_{r,cr} (T, r = 0 ) & = & - { { \rho_0 f' (0)} \over {T_0} }
\; { { 1 - \alpha_r} \over { 2 - \alpha_r} } \; { { d+z } \over z} \;
\nonumber \\
& & \times \left( { T \over {T_0} } \right)^{ 
( d ( 1 - \alpha_r ) - z ) / ( z ( 2 - \alpha_r) ) } \;,
\nonumber \\
\chi_{r,cr} (T, r = 0 ) & = & - \rho_0 f''( 0 ) 
\left( { T \over {T_0} } \right)^{- ( ( d + z ) \alpha_r) 
/ ( z ( 2 - \alpha_r))} \;.
\label{critexpr}
\end{eqnarray}
On the other hand, in the massless phase characterized by a specific
heat exponent $y_0$, at low temperatures in the vicinity of the QPT,
\begin{eqnarray}
C_{cr} (T \rightarrow 0, r) & = & { { \rho_0 c} \over {T_0} } 
\; y_0 ( y_0 + 1 ) \; r^{ ( 2 - \alpha_r) ( d - y_0 z) /(d +z)} 
\nonumber \\  
& & \times \left( { T \over {T_0} } \right)^{y_0} \;,
\nonumber \\
\eta_{r,cr} (T \rightarrow 0, r) & = & - { { \rho_0 c} \over {T_0} } 
 \; { {( d - y_0 z)} \over { d + z } } \; (y_0 + 1) \;
( 2 - \alpha_r)  
\nonumber \\
& & \times r^{ ( 2 - \alpha_r) ( d - y_0 z) /(d +z) - 1}
\left( { T \over {T_0} } \right)^{y_0} \;,
\nonumber \\
\chi_{r,cr} (T \rightarrow 0, r) & = & - \rho_0 \tilde{f} (0)
( d + z - 1) ( 2 - \alpha_r) r^{ - \alpha_r}
\nonumber \\ 
& & -
c \rho_0 { {( d - y_0 z)} \over {d+z} } 
( ( 2 - \alpha_r) \left( {{ d - y_0 z} \over { d+z} } 
\right) - 1 )
\nonumber \\ 
& & \times r^{ ( 2 - \alpha_r) ( d - y_0 z) /(d +z) - 2}
\left( { T \over {T_0} } \right)^{y_0 + 1} \;.
\label{maslessexp}
\end{eqnarray}
From the above equations one directly infers the main results from
Rosch-Si\cite{roschsi}: the `Gr\"uneisen ratio' $\Gamma_r = \eta_r / C \sim
T^{-y_r/z}$ in the quantum critical state while in the massless 
phase it becomes exactly $ ( d - y_0 z) / (y_0 y_r) r^{-1}$, i.e.
it acquires an universal amplitude expressed entirely in terms of 
the exponents. The significance of the coupling
constant susceptibility $\chi_r$ is immediately clear from
 Eq.'s (\ref{critexpr},\ref{maslessexp}). Its temperature
dependence reveals that it is more singular than $\eta_r$, which is
in turn more singular than $C$. In addition, its temperature
independent part  diverges in the approach to the
critical point with the exponent $\alpha_r$, in direct
analogy with the divergence of the specific heat with $\alpha$
in the approach to a thermal phase transition.

Let us now apply the  above scaling laws to the specific context  
encountered in the cuprates. By 
restricting ourselves to thermodynamics we have to assume
very little in addition to Eq. (\ref{freenscal}): (i) In the
cuprates the relevant zero-temperature direction is the electron
density varied by the doping $p$. The
reduced coupling constant corresponds therefore with 
$ x = ( p - p_c) / p_c $ (ii) Recently, evidences have been
accumulating showing that the overdoped state is a Fermi-liquid, 
characterized by $y_0 = 1$\cite{campuzano,tagaki,hussey}.
(iii) We rely on the specific heat as measured 
by  Loram, Tallon and coworkers\cite{loram}.  
Since the superconductivity appears to hide
the critical behavior, the regime of interest is at high temperature. 

Given the assumption that electron density is the zero temperature control 
parameter it follows from elementary thermodynamics that the quantities
$\eta_r$ and $\chi_r$ relate to $\mu$,  
\begin{eqnarray}
\eta_{cr, x}  & = &  { {\partial S_{cr}} \over {\partial x} } |_{\mu}
 =   - { {\partial \mu} \over {\partial T} } |_{x} \;,
\nonumber \\
\chi_{cr, x} & = & { {\partial^2 F_{cr}} \over {\partial x^2} } |_{\mu}
 =  { {\partial \mu} \over {\partial x} } = { 1 \over {n^2 \kappa}} \;,
\label{muTn}
\end{eqnarray}
where $\kappa$ is just the electronic compressibility and $n$ the
total electron density. Notice that
when pressure is the control parameter, $\chi \sim 
 \partial^2 F / \partial p^2  \sim  \partial V / \partial p $ refers
to the total compressibility. 

Let us now turn to the measured electronic specific heat of the
cuprates\cite{loram}. In fact, the remarkable property of the measured 
specific heat is
its uninteresting appearance. In the overdoped regime it is indistinguishable 
from the specific heat of a conventional BCS superconductor. At high 
temperatures, $C = \gamma T$ with a temperature independent $\gamma$
like in a Fermi-liquid, and at the superconducting transition the
specific heat shows a BCS-like anomaly. Upon decreasing doping, all what
happens is that the pseudogap scale manifests itself quite clearly in
the form of a decreasing $\gamma$, a fact exploited by Loram {\em et al.}
to study the doping dependence of the pseudo gap temperature $T_{PG}$. Above
$T_{PG}$ $\gamma$ is temperature independent  and connected smoothly with the
specific heat in the 
overdoped regime, showing no noticable doping dependence. 

It seems
to be a reflex to assume that the `metallic' appearance of the 
$\gamma$ above $T_{PG}$ is just revealing that a Fermi-liquid state 
is re-established at high temperatures, but this is actually
quite unreasonable. Recently, evidences has been
accumulating that on the overdoped side a `coherence' crossover
occurs: one can identify a temperature $T_{coh}$ below which transport shows
Fermi-liquid signatures\cite{tagaki,hussey} while photoemission reveals 
that the quasiparticles become underdamped\cite{campuzano}. 
$T_{coh}$ emerges at optimal doping and increases 
with increasing doping in the overdoped regime. It is no wonder that the
low temperature specific heat in this Fermi-liquid regime is conventional,
but why is it so that it remains conventional  above $T_{coh}$? 
Stronger, why is it unaltered  at temperatures $> T_{SG}$ even in the
strongly underdoped regime?

Let us reconsider the scaling of the specific heat in the QC regime,
Eq. (\ref{critexpr}). The remarkable fact is that its
temperature dependence is predicted to be uninteresting!  
Its temperature exponent
is just given by the ratio of the number of space- ($d$) and effective
time ($z$) dimensions. 
In the quantum critical regime of the cuprates $C \sim T$ and this
means that $d=z$, the number of space dimensions equals the number of
time dimensions! At these high temperatures, it seems reasonable to
assume that $d=2$, with the implication that $ z = 2$, signalling
diffusion.    

There is a non-trivial consistency with the observation that the
specific heat is not sensitive to the crossover from the
quantum critical- to the Fermi-liquid regime at $T_{coh}$.
From Eq. (\ref{maslessexp}) it follows that the specific 
heat in a massless state knows about the proximity of the QPT via 
the factor $r^{ ( 2 - \alpha_r) ( d - y_0 z) /(d +z)}$, governing
the divergence of the quasiparticle mass. The exponent
contains the combination of the dimensions $d - y_0 z$ 
and when $d = z$ and $y_0 = 1$ as in the Fermi-liquid the
exponent vanishes  and the specific heat becomes insensitive 
to the zero temperature singularity! 
The specific heat is expected to be just the same at all
temperatures and dopings as long as $T > T_{SG}$ despite the
fact that other properties demonstrate large scale changes 
in the physics.

To further stress this point, let us consider what happens
in the pseudo-gap regime $T < T_{SG}$. The measured specific
heat shows that in between the superconducting $T_c$ and
$T_{SG}$ $C \sim T^2$ and thermodynamically it can be viewed
as a `nodal liquid' characterized by $y_0 =2$. Insisting that
$d = z$  it follows from Eq. (\ref{maslessexp}) that 
$C \sim r^{ - ( 2 - \alpha_r) /2} T^2$.  From Eq.(\ref{freeenwork})
it follows immediately that the pseudogap scale
$T_{SG} \sim r^{ z /y_r} = r^{( 2 - \alpha_r) /2}$; it just means
that $C \sim T^2 / T_{SG}$ which is consistent with experiment.
Notice that this would fail when $ d \neq z$.

Because $\alpha_r$ is expected to be small, $T_{SG}$ is expected to
be weakly sublinear in $x$ when $d=z$.
In a recent paper\cite{zntallon}, the behavior of $T_{SG}$ for small
$x$ has been determined in 123 samples where the superconductivity has been
surpressed by $Zn$ doping. $T_{SG}$ turns out to be indeed weakly
sublinear in $x$, suggesting that $\alpha_r$ is in the range $0.2-0.3$, i.e.
a reasonable value for a strongly interacting unstable fixed point.   
 
Up to this point we have presented the case that if a QPT is present at
optimal doping,  the quantum singularity is largely hidden from
the specific heat for specific reasons ($d=z$, the Fermi-liquid).
 To establish the presence of this singularity one
has to look elsewhere and the remedy is obvious: the thermodynamic
potential. Assuming $d=z$ one finds an interesting collection of
scaling behaviors for $\partial \mu / \partial T$  and the
inverse compressibility $\chi_x$. 

Omitting non-universal factors
and including the specific heat for completeness, these become
in the quantum-critical regime,
\begin{equation}
C_{cr}  \sim  T; \; \;
{ {\partial \mu} \over {\partial T} }_{cr}  \sim  - T^{- { {\alpha_r} \over 
{2 - \alpha_r} } }; \; \;
\chi_{cr, x}   \sim   - T^{- { {2\alpha_r} \over  {2 - \alpha_r} } } \;.
\label{critdz}
\end{equation}
Hence, by measuring the temperature dependences of the chemical potential
and the compressibility in the {\em high temperature} quantum critical
regime one  obtains directly the `quantum alpha' characterizing the
nature of the quantum singularity. Notice that  the 
incompressibility should be precisely twice as singular as the temperature
derivative of $\mu$.

The Fermi-liquid regime ($y_0 = 1$) is not particularly revealing,
\begin{equation}
C_{FL,cr}  \sim  T; \; \; 
{ {\partial \mu} \over {\partial T} }_{FL,cr} =  0; \; \;
\chi_{FL, cr, x}   \sim  x^{-\alpha_r} \;.
\label{critFL}
\end{equation}
The critical part of $\partial \mu / \partial T$ vanishes because
the prefactor contains $d- y_0 z$ as does the temperature dependent
part of $\chi_n$. Only the temperature independent part of the
inverse compressibility reveals directly the quantum singularity.

In the pseudogap regime ($y_0 = 2$)  this changes drastically. 
Parametrizing matters in terms of the pseudogap scale
$T_{SG} (n) \sim x^{ (2 - \alpha_r) /  2 }$,
\begin{eqnarray}
C_{SG,cr}  & \sim &  { { T^2 } \over T_{SG} (x) }; \; \; 
{ {\partial \mu} \over {\partial T} }_{SG,cr}  =   { { C_{SG,cr}}
\over x}; \nonumber \\
\chi_{SG, cr, x} &   \sim  & A x^{-\alpha_r} - B { {C_{SG,cr} T } 
\over {x^2} } \;.
\label{critSG}
\end{eqnarray}
The second and third line  reflect the workings
of the `generalized Grueneisen parameters'  as realized by
Rosch-Si\cite{roschsi}. 
$\partial \mu / \partial T$ is clearly `one order
more singular' in $n$ than the specific heat, but the
temperature dependent part of the incompressibility  is actually
`twice as singular'. 
  
Summarizing, using elementary power counting arguments we have 
discovered an emperical strategy which should make it possible 
to decide if the `quantum criticality' of the cuprates has to
do with universality. We have found that the temperature- and
density dependences of the chemical potential in the various
regimes should obey {\em six} scaling laws, which are all
governed by a single fundamental scaling dimension ($\alpha_r$).
As an input, we have used the specific heat data to argue
that the effective number of space ($d$) and time ($z$) dimensions   
characterizing the critical state have to be the same.   

We are not aware of chemical potential- and electronic 
compressibility measurements of the cuprates having the
required accuracy. However, this does not appear
to represent a problem of principle. In the experimental literature
one finds a variety of methods to measure these 
quantities\cite{fujimori,polish}, a prime example being the 
vibrating Kelvin probe method allowing for high accuracy measurements
of the chemical potential, used by van der Marel and coworkers 
sometime ago to determine the density dependence of the superconducting
$T_c$\cite{rietveld}. We suggest to use these experimental
methods to establish once and for all the presence or absence of a
genuine quantum phase transition in the cuprates. 

We notice that our
scaling relations might also be put to the test in the context
of the metal-insulator transition in the two dimensional electron gas.
Using various ingenious techniques\cite{ilani,dultz}, 
the electronic compressibility has
been measured in the proximity of this quantum phase transition and 
it would be highly interesting to focus on the temperature dependence
of the chemical potential. In a forthcoming publication we will address
this problem in more detail.

{\em Acknowledgements}
We acknowledge helpful discussions with  J.M.J. van Leeuwen, J.W. Loram,
D. van der Marel, A. Rosch, S. Sachdev, and J.L. Tallon.

\end{multicols}
\end{document}